\documentclass[]{pasj01}
\draft

\begin{document} 
\Received{}
\Accepted{}

\title{Revised Wavelength and Spectral Response Calibrations for AKARI Near-Infrared Grism Spectroscopy: Cryogenic Phase}

\author{Shunsuke \textsc{baba}\altaffilmark{1,2}}
\altaffiltext{1}{Institute of Space and Astronautical Science, Japan Aerospace Exploration Agency, 3-1-1 Yoshinodai, Chuo-ku, Sagamihara, Kanagawa 252-5210, Japan}
\altaffiltext{2}{Department of Physics, Graduate School of Science, The University of Tokyo, 7-3-1 Hongo, Bunkyo-ku, Tokyo 113-0033, Japan}
\email{s-baba@ir.isas.jaxa.jp}

\author{Takao \textsc{nakagawa}\altaffilmark{1}}

\author{Mai \textsc{shirahata}\altaffilmark{3}}
\altaffiltext{3}{Frontier Research Institute for Interdisciplinary Science, Tohoku University, 6-3 Aramaki Aza-Aoba, Aoba-ku, Sendai, Miyagi 980-8578, Japan}

\author{Naoki \textsc{isobe}\altaffilmark{1}}

\author{Fumihiko \textsc{usui}\altaffilmark{4}}
\altaffiltext{4}{Department of Astronomy, Graduate School of Science, The University of Tokyo, 7-3-1 Hongo, Bunkyo-ku, Tokyo 113-0033, Japan}

\author{Youichi \textsc{ohyama}\altaffilmark{5}}
\altaffiltext{5}{Institute of Astronomy and Astrophysics, Academia Sinica, P.O. Box 23-141, Taipei 10617, Taiwan}

\author{Takashi \textsc{onaka}\altaffilmark{4}}

\author{Kenichi \textsc{yano}\altaffilmark{1,2}}

\author{Chihiro \textsc{kochi}\altaffilmark{1,2}}


\maketitle

\begin{abstract}
We perform revised spectral calibrations for the AKARI near-infrared grism to quantitatively correct for the effect of the wavelength-dependent refractive index.
The near-infrared grism covering the wavelength range of 2.5--5.0\,\micron\ with a spectral resolving power of 120 at 3.6\,\micron, is found to be contaminated by second-order light at wavelengths longer than 4.9\,\micron\ which is especially serious for red objects.
First, we present the wavelength calibration considering the refractive index of the grism as a function of the wavelength for the first time.
We find that the previous solution is positively shifted by up to 0.01\,\micron\ compared with the revised wavelengths at 2.5--5.0\,\micron.
In addition, we demonstrate that second-order contamination occurs even with a perfect order-sorting filter owing to the wavelength dependence of the refractive index.
Second, the spectral responses of the system from the first- and second-order light are simultaneously obtained from two types of standard objects with different colors.
The response from the second-order light suggests leakage of the order-sorting filter below 2.5\,\micron.
The relations between the output of the detector and the intensities of the first- and second-order light are formalized by a 
matrix equation that combines the two orders.
The removal of the contaminating second-order light can be achieved by solving the matrix equation.
The new calibration extends the available spectral coverage of the grism mode from 4.9\,\micron\ up to 5.0\,\micron.
The revision can be used to study spectral features falling in these extended wavelengths, e.g., the carbon monoxide fundamental ro-vibrational absorption within nearby active galactic nuclei.
\end{abstract}

\section{Introduction}

AKARI is the Japanese infrared satellite launched on February 21, 2006 \citep{MurakamiEtAl_2007}.
The satellite is equipped with a 68.5-cm telescope and two focal plane instruments.
These assemblies are cryogenically cooled by liquid helium and mechanical cryocoolers \citep{NakagawaEtAl_2007}.
One of the focal plane instruments is the Infrared Camera (IRC; \cite{OnakaEtAl_2007}), which is dedicated to imaging and spectroscopic observations in the near- and mid-infrared wavelength regions.
The main purpose of the AKARI mission is to carry out an all-sky survey.
In addition to the all-sky survey observations, a number of pointing observations have been carried out with the IRC.

The IRC comprises three channels, each of which covers a different wavelength range:
NIR (1.8--5.5\,\micron), MIR-S (4.6--13.4\,\micron), and MIR-L (12.6--26.5\,\micron).
A prism (NP) and a grism (NG) are equipped with the NIR channel for spectroscopy.
The grism mode covers the 2.5--5.0\,\micron\ range with a spectral resolving power ($R\equiv\lambda/\Delta\lambda$) of 120 at 3.6\,\micron, whereas the prism mode is 1.8--5.5\,\micron\ with lower resolution of $R=19$ at 3.5\,\micron\ \citep{OhyamaEtAl_2007e}.
The NIR spectroscopy is one of the unique capabilities of AKARI because continuous coverage in this wavelength range cannot be achieved by ground-based telescopes owing to atmospheric absorption.
The Infrared Space Observatory (ISO; \cite{KesslerEtAl_1996}) is also capable of near-infrared spectroscopy but its observations are limited to bright sources.
The sensitivities of ISO and AKARI are on the order of a few Jy and a few mJy, respectively.
The Infrared Spectrometer (IRS) onboard the Spitzer space telescope observes wavelengths longer than 5.2\,\micron\ \citep{HouckEtAl_2004}.

The IRC team has developed and distributed the IDL-based data reduction package ``IRC Spectroscopy Toolkit'' at the AKARI website\footnote{\texttt{http://www.ir.isas.jaxa.jp/AKARI/Observation/support/IRC/}}.
The toolkit extracts a spectral image for each source after the basic data reduction (dark subtraction, linearity correction, flat fielding, stacking, etc.).
For each spectral image, the toolkit converts the pixel positions along the dispersion direction to wavelengths and divides the outputs by the spectral response (the output per the unit flux density at each wavelength).
The one-dimensional calibrated spectrum is obtained by summing up the image along the spatial axis \citep{OhyamaEtAl_2007e}. 

``AKARI IRC Data User Manual for Post-Helium (Phase 3) Mission Version 1.1''\footnote{\S 6.9.7. \texttt{http://www.ir.isas.jaxa.jp/AKARI/Observation/support/IRC/IDUM/IRC\_IDUM\_P3\_1.1.pdf}} states that the second-order light diffracted by the grism overlaps with the first-order light of wavelengths longer than 4.9\,\micron.

Previous spectral response calibrations of the grism mode performed by \citet{OhyamaEtAl_2007e} and T. Shimonishi\footnote{see the change log of the IRC Spectroscopy Toolkit version 20130813 and \citet{ShimonishiEtAl_2013}} are based on observations of A- and K-type standard stars.
These stars have blue spectra; in other words, their fluxes are larger at shorter wavelengths.
Consequently, the intensity of the contaminating second-order light increases relative to that of the first-order light at longer wavelengths. 
Observed signals for 4.9--5.0\,\micron\ are therefore vulnerable to second-order light and, hence, the spectral response in this wavelength range produces erroneous flux values particularly for red objects, which have less contamination from the second-order light.
Previous flux calibrations neglected this contamination effect.

The wavelength range contaminated by the second-order light is important for some features, such as the carbon monoxide fundamental ro-vibrational absorption bands (around at 4.7\,\micron).
For instance, such absorption bands can be used to probe the physical conditions of molecular clouds surrounding active galactic nuclei (AGNs), as investigated by, e.g., \citet{SpoonEtAl_2004b} and \citet{ShirahataEtAl_2013}.
It is difficult to analyze the absorption from the grism mode spectra unless the contamination is corrected properly.

In this paper, we quantify and correct for the contamination of the second-order light in the IRC grism spectroscopy.
In section \ref{sec. design}, the design of the NIR channel and that of the grism are described.
We revise the wavelength calibration of the grism mode more precisely and discuss the spectral responses from the first- and second-order light in section \ref{sec. new_cal}.
The usefulness of the proposed flux calibration that corrects for the effect of the second-order light is discussed in section \ref{sec. flux-cal}.
Finally, we summarize the calibration results in section \ref{sec. summary}.

\section{Design of the IRC NIR Channel and the Grism}
\label{sec. design}

A grism is a prism combined with a transmission diffraction grating designed so that a ray of specific wavelength passes straight through.
With the help of the diffraction grating, the grism has higher spectral resolving power than the prism, but at the same time, it produces higher order interference.
In simple grating spectroscopy, the observing wavelength range is limited within an octave.
An order-sorting filter is generally used to prevent overlap between light of different orders.
The situation is more complex in the case of grism spectroscopy.
The optical path difference within a grism depends on the refractive index of its material.
Owing to the wavelength dependence of the refractive index, the second-order light can contaminate the first-order spectrum even if the order-sorting filter works perfectly.
We therefore review the design of the IRC NIR channel and the grism considering this effect.

\begin{figure}
  \begin{center}
    \includegraphics[width=16cm]{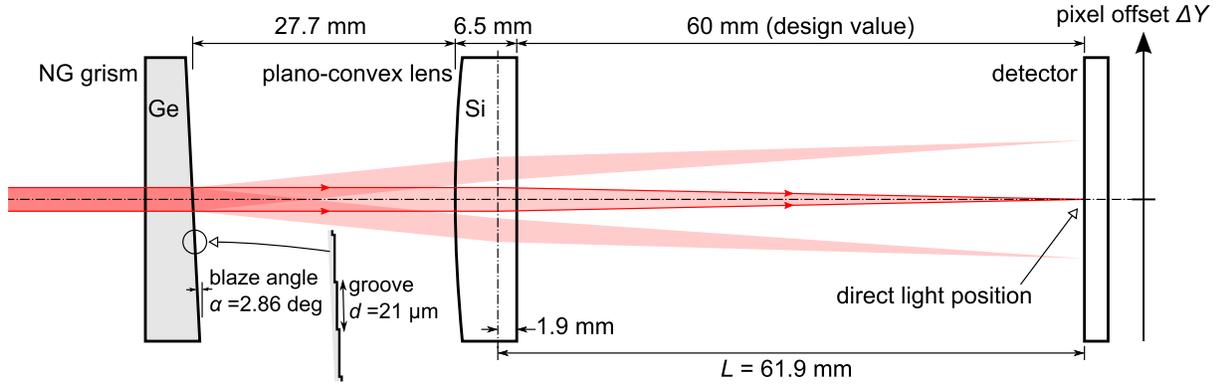}
  \end{center}
  \caption{Schematic of the NIR channel. See the text for explanation.}
  \label{fig. design_NG}
\end{figure}

According to the ASTRO-F interim report volume 2 \citep{ASTRO-F_2002}, the light that enters the NIR channel is first collimated by three lenses, diffracted by the grism mounted on the filter wheel, and focused by the plano-convex camera lens on the detector array.
Figure \ref{fig. design_NG} shows a schematic of the NIR channel and key parameters.
\citet{OnakaEtAl_2007} give a more detailed layout including the fore-optics.
The spacing between the grism and the plano-convex lens is 27.7\,mm.
This lens is 6.5-mm thick and made of silicon.
Since the refractive index of silicon is 3.4 in the wavelength range of the grism mode \citep{FreyEtAl_2006}, the second principal plane of the plano-convex lens is located $6.5/3.4=1.9\,\mathrm{mm}$ in front of its rear surface.
The designed interval between the lens and the focal plane is 60\,mm.
Note that the actual in-flight interval between them may not be equal to this value because the focal position is optimized for low-temperature operation by slightly rearranging the optical elements.
The report does not provide the in-flight value.
Thus, we assume the design value of 60\,mm for the interval as a first approximation.

IR light of wavelength $\lambda$ interferes constructively with the grism if the emitting angle $\theta$ satisfies
\begin{equation}
  m \lambda = d \left[ n \sin\alpha - \sin (\alpha - \theta) \right],
  \label{eq. interfere}
\end{equation}
where $m$ is an integer that denotes the order number, $\alpha$ is the blaze angle, $d$ is the groove spacing, and $n$ is the refractive index of the material of the grism.
The grism is made of germanium (Ge).
The design value of the blaze angle is $\alpha = 2.86^\circ$ and the groove spacing is $d = 21\,\micron$ \citep{OhyamaEtAl_2007e}.
The interval from the second principal plane of the lens to the detector is $L=61.9\,\mathrm{mm}$ (figure \ref{fig. design_NG}); thus, IR rays of the emitting angle $\theta$ converges at a point separated by $L\tan\theta$ from the direct light point on the detector.
Dividing the separation by the pixel pitch of the detector $p=30\,\micron$, the pixel offset from the direct light position $\Delta Y$ (in units of pixel number) is represented as
\begin{equation}
  \Delta Y= (L/p)\tan\theta.
  \label{eq. Delta-Y}
\end{equation}
Consequently, the relation between wavelength $\lambda$ and pixel offset $\Delta Y$ is modeled by equations (\ref{eq. interfere}) and (\ref{eq. Delta-Y}).
When $\theta$ is small, equations (\ref{eq. interfere}) and (\ref{eq. Delta-Y}) are simplified to
\begin{equation}
  \Delta Y = \frac{L}{p} \left[ \frac{m\lambda}{d} - (n - 1) \alpha \right].
  \label{eq. linear}
\end{equation}

If the refractive index $n$ has no wavelength dependence, the pixel position linearly relates with the wavelength; hence, the first-order 5.0\,\micron\ and the second-order 2.5\,\micron\ correspond to the same position.
The current toolkit uses a linear wavelength calibration based on this assumption, and it is expressed as 
\begin{equation}
  \lambda_\mathrm{tool} [\micron] = 0.00967625 \times \Delta Y [\mathrm{pix}] + 3.12121.
  \label{eq. tool}
\end{equation}
The front surface of the grism has multi-layer coating that cuts off the radiation with wavelengths shorter than 2.5\,\micron\ to avoid the contamination of the second-order light as an order-sorting filter.
However, the refractive index of Ge changes by up to 1\% as a function of wavelength between 2.5 and 5.0\,\micron\ \citep{FreyEtAl_2006}.
Moreover, the cut-off coating is not perfect and, in practice, there is leakage.
Contamination by second-order light can occur even at wavelengths shorter than 5.0\,\micron.

\begin{figure}
  \begin{center}
    \includegraphics[width=8cm]{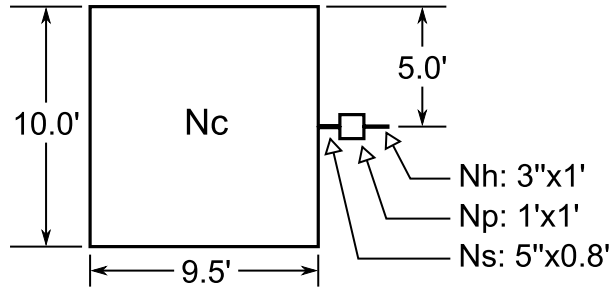}
  \end{center}
  \caption{Field of view of the NIR channel.}
  \label{fig. FOV}
\end{figure}

The field-of-view of the NIR channel consists of four sections, each of which has a respective observational purpose.
These sections are composed of two slits and two square-shaped apertures, which are shown in figure \ref{fig. FOV}.
The wider slit is named Ns and the narrower one is named Nh.
The width of the Ns and Nh slits are $5''$ and $3''$, respectively.
These slits are designed for observing extended sources.
In the larger aperture (Nc), multi-object slit-less spectroscopy and imaging observations can be performed.
The smaller aperture (Np) is equipped to observe a point source avoiding overlaps with other sources.
This aperture has size of $1'\times 1'$ \citep{OnakaEtAl_2007, OhyamaEtAl_2007e}.

The design of the grism mode does not precisely consider the wavelength dependence of the refractive index of the grism.
The wavelength dependence can cause the contamination of the second-order light and must be considered in the flux calibration.
The refractive index also depends on the temperature as well as the wavelength.
In this paper, we focus on observations before the exhaustion of liquid helium (Phases 1 and 2), when the temperature of the NIR channel was stable.
Similar calibrations for the post-helium phase (Phase 3) will be conducted in the future.

\section{New Calibration Method for the IRC Grism Spectroscopy}
\label{sec. new_cal}

\subsection{Wavelength Calibration}
\label{subsec. wave-cal}

In this section, based on observations of objects that show several emission lines, we revise the relation between the pixel offset and the wavelength for the grism mode using the equations described in the previous section and considering the refractive index as a function of the wavelength.
In addition to the relation for the first-order light, we also calculate the relation for the second-order light to identify the spectral range where the contamination occurs.

The wavelength-dependent refractive index $n(\lambda)$ of Ge was measured at cryogenic temperatures by \citet{FreyEtAl_2006}, and they give it as a function of wavelength and temperature in the range of 1.9--5.5\,\micron\ and 20--300\,K.
The NIR channel was cooled down to $\sim 6\,\mathrm{K}$ during Phases 1 and 2 \citep{NakagawaEtAl_2007}.
Extrapolation of the function provided by \citet{FreyEtAl_2006} to lower temperatures suggests that the temperature dependence of the refractive index between 6 and 20\,K is negligibly small ($\sim0.01\%$).
We use $n(\lambda)$ at 20\,K as the operating refractive index of the grism in the following analysis.

\begin{figure}
  \begin{center}
    \includegraphics[width=8cm]{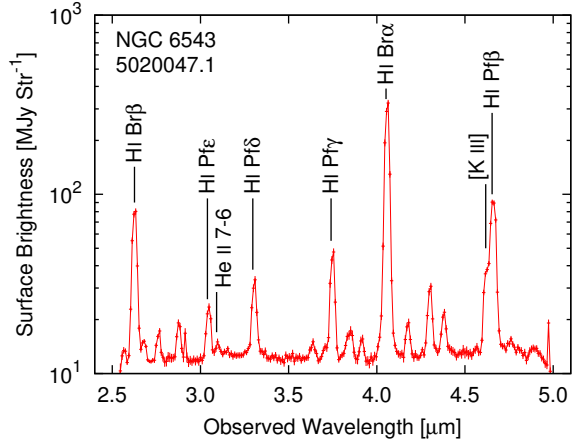}
  \end{center}
  \caption{2.5--5.0\,\micron\ spectra of NGC 6543 observed in the grism mode (observation ID: 5020047.1).}
  \label{fig. spec_NGC6543}
\end{figure}

The previous calibration expressed in equation (\ref{eq. tool}) is made based on observations of the recombination lines of the bright planetary nebula NGC 6543 \citep{OhyamaEtAl_2007e}.
To examine the effect of the wavelength dependence of the refractive index, we revisit the two observations of NGC 6543 (Observation IDs 5020047.1 and 5020048.1).
Both observations were carried out on April 29, 2006 using the Ns slit.
We follow Yano et al. (in preparation) for the data reduction and the error estimate.
One of the reduced spectra is shown in figure \ref{fig. spec_NGC6543}.
We fit a Gaussian on a linear continuum to each recombination line excepting H\emissiontype{I} Pf$\beta$ and H\emissiontype{I} Pf$\varepsilon$.
H\emissiontype{I} Pf$\beta$ is fitted together with its immediate neighbor [K\emissiontype{III}] by two Gaussians on a linear continuum.
Similarly, H\emissiontype{I} Pf$\varepsilon$ is fitted with He\emissiontype{II} 7--6.
The fitted central wavelengths $\lambda_\mathrm{tool}$ are tabulated in table \ref{tab. lines} with theoretical values $\lambda_\mathrm{true}$ taken from the ISO line list\footnote{\texttt{http://www.mpe.mpg.de/ir/ISO/linelists/Hydrogenic.html}}.
The difference between $\lambda_\mathrm{tool}$ and $\lambda_\mathrm{true}$ relative to $\lambda_\mathrm{true}$ corresponds to velocity of 450\,km\,s${}^{-1}$ on average.
The velocity of NGC 6543 relative to the local standard of rest is $V_\mathrm{LSR}=-51\,\mathrm{km\,s}^{-1}$  \citep{SchneiderEtAl_1983}.
The Doppler shift arising from $V_\mathrm{LSR}$ is smaller than the difference between $\lambda_\mathrm{tool}$ and $\lambda_\mathrm{true}$ by an order of magnitude.
For the two observations, the orbital velocity of the satellite projected to the line-of-sight is 0.7\,km\,s${}^{-1}$.
The Doppler shift by the satellite's motion is also smaller than the difference $\lambda_\mathrm{tool}-\lambda_\mathrm{true}$ by a few orders of magnitude.
We here ignore any Doppler effects.

\begin{table}
  \tbl{Fitted recombination lines.}{%
    \begin{tabular}{cccc}
      \hline
      line                              & $\lambda_\mathrm{true}$ (\micron)\footnotemark[$*$] 
                                                  & \multicolumn{2}{c}{$\lambda_\mathrm{tool}$ (\micron)\footnotemark[$\dagger$]} \\
                                                  \cline{3-4}
                                        &         & 5020047.1            & 5020048.1                      \\
      \hline
      H\emissiontype{I} Br$\beta$       & 2.62587 & $2.62629\pm 0.00087$ & $2.62559\pm 0.00044$ \\
      H\emissiontype{I} Pf$\varepsilon$ & 3.03920 & $3.04600\pm 0.00059$ & $3.04581\pm 0.00049$ \\
      H\emissiontype{I} Pf$\delta$      & 3.29699 & $3.30523\pm 0.00054$ & $3.30501\pm 0.00065$ \\
      H\emissiontype{I} Pf$\gamma$      & 3.74056 & $3.74909\pm 0.00066$ & $3.74914\pm 0.00049$ \\
      H\emissiontype{I} Br$\alpha$      & 4.05226 & $4.05811\pm 0.00035$ & $4.05823\pm 0.00045$ \\
      H\emissiontype{I} Pf$\beta$       & 4.65378 & $4.65898\pm 0.00067$ & $4.65886\pm 0.00086$ \\
      \hline
    \end{tabular}
  }
  \label{tab. lines}
  \begin{tabnote}
    \footnotemark[$*$] The theoretical wavelength of each line taken from the ISO line list.\\
    \footnotemark[$\dagger$] The wavelength of each line in the spectra is processed using the toolkit (figure \ref{fig. spec_NGC6543}).
  \end{tabnote}
\end{table}

\begin{figure}
  \begin{center}
    \includegraphics[width=8cm]{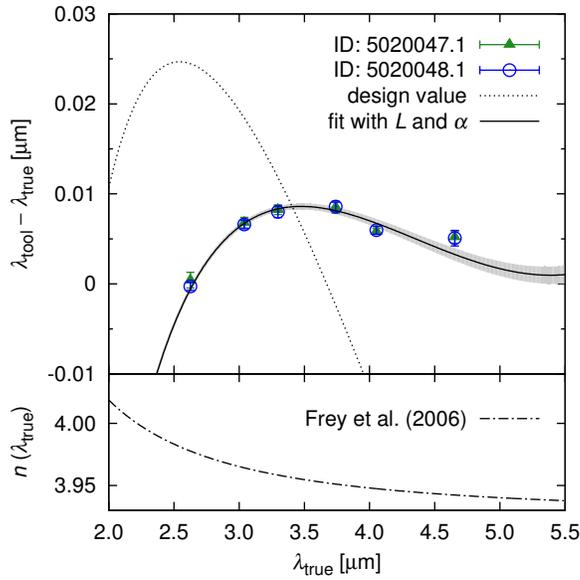}
  \end{center}
  \caption{Top: The points indicate the differences between the tabulated wavelengths in table \ref{tab. lines}. The dotted line represents the relation predicted from equations (\ref{eq. interfere}), (\ref{eq. Delta-Y}), and (\ref{eq. tool}), and the refractive index shown in the bottom panel. The measured values are plotted as triangles and open circles. The solid line is the curve obtained by fitting the dotted line to the points with variable parameters $L$ and $\alpha$. The gray-shaded area denotes the one-sigma uncertainty of the best fit. Bottom: refractive index of Ge at 20\,K \citep{FreyEtAl_2006}.}
  \label{fig. tool-true}
\end{figure}

The differences between $\lambda_\mathrm{tool}$ and $\lambda_\mathrm{true}$ can be explained by the wavelength dependence of $n(\lambda)$.
From equations (\ref{eq. interfere}), (\ref{eq. Delta-Y}), and (\ref{eq. tool}), and $n(\lambda)$, the relation between the difference of $\lambda_\mathrm{tool}-\lambda_\mathrm{true}$ and $\lambda_\mathrm{true}$ can be estimated.
Here, $\lambda$ in equation (\ref{eq. interfere}) is assumed as $\lambda_\mathrm{true}$.
The estimates are denoted by the dotted line in figure \ref{fig. tool-true} and fail to reproduce the measured differences.
The failure may be attributed to the possibility that the assumed $L=61.9\,\mathrm{mm}$ differs from the in-flight value.
Thus, we take $L$ and the blaze angle $\alpha$ as variables and fit the curve to the differences.
The obtained best fit is denoted by the solid line in figure \ref{fig. tool-true}.
The fitted parameters are $L=63.92\pm 0.03\,\mathrm{mm}$ and $\alpha=2.8690\pm 0.0003^\circ$.
The change in $L$ is sufficiently small to be interpreted as the result of the adjustment of the focal length described in the previous section, and that in $\alpha$ is within the fabrication error.
Hereafter, we adopt the best-fit parameters in the following calibrations.

\begin{figure}
  \begin{center}
    \includegraphics[width=8cm]{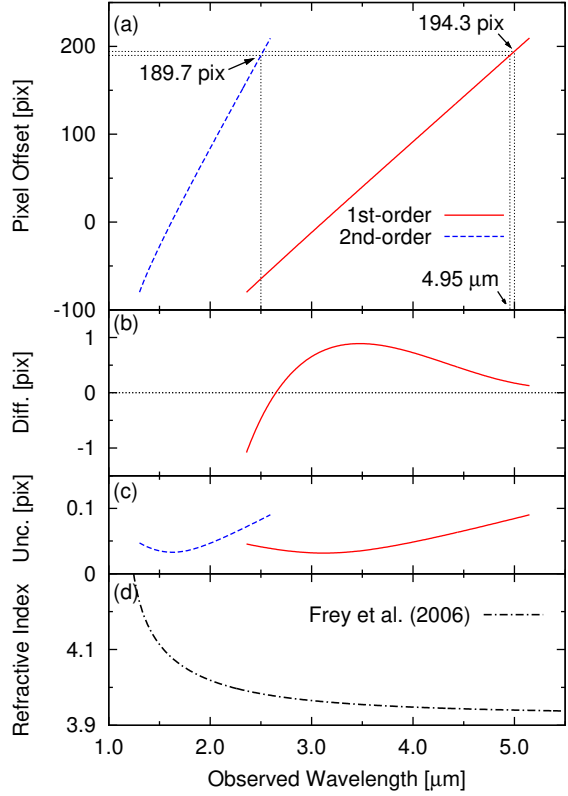}
  \end{center}
  \caption{(a) The relation between the pixel offset and the wavelength for the first- and second-order light calculated with the wavelength-dependent refractive index of Ge. (b) Difference of the pixel offset for the first-order light between the new and old wavelength calibration. The pixel offset of the present wavelength calibration (equation \ref{eq. tool}) is subtracted from that shown in the top panel. (c) One-sigma uncertainty of the pixel offset. (d) Refractive index of Ge at 20\,K \citep{FreyEtAl_2006}, which is the same as the bottom panel of figure \ref{fig. tool-true} but is also shown in shorter wavelengths.}
  \label{fig. wavelength_vs_pos}
\end{figure}

Using the best fit, the relation between $\Delta Y$ and $\lambda$ is calculated for both the first- and the second-order light.
The obtained relations are shown in figure \ref{fig. wavelength_vs_pos}(a).
We denote the wavelength of the first-order light as $\lambda^{(1)}$ and that of the second-order light as $\lambda^{(2)}$.
$\lambda^{(1)}=5.00\,\micron$ goes to $\Delta Y = 194.3\,\mathrm{pix}$, whereas $\lambda^{(2)}=2.50\,\micron$ goes to $\Delta Y = 189.7\,\mathrm{pix}$.
The difference of 4.6\,pix stems from the wavelength dependence of the refractive index of Ge.
The refractive index $n$ is 3.940 at 5.00\,\micron\ and 3.983 at 2.50\,\micron\ (figure \ref{fig. wavelength_vs_pos}(d)).
According to equation (\ref{eq. tool}), $\Delta n$ yields the difference of the pixel offset $(L/p) \Delta n \alpha$.
$\Delta Y = 189.7\,\mathrm{pix}$ corresponds to $\lambda^{(1)}=4.95\,\micron$.
Hence, even if the order-sorting filter of the grism perfectly cuts the radiation of $\lambda^{(2)}<2.50\,\micron$, the second-order light contaminates the first-order spectra at $\lambda^{(1)}=4.95$--$5.00\,\micron$.
The derived pixel offset for the first-order light differs from that of the present wavelength calibration by up to 0.8\,pixel at 3.5\,\micron\ (figure \ref{fig. wavelength_vs_pos}(b)).
This is a direct consequence of the fact that the observed wavelength difference $\lambda_\mathrm{tool}-\lambda_\mathrm{true}$ reaches the maximum 0.008\,\micron\ at $\lambda_\mathrm{true}=3.5\,\micron$ (see figure \ref{fig. tool-true}).
This difference is smaller than the shift of the positions for the first- and second-order light at about $\Delta Y \sim 190\,\mathrm{pix}$ discussed above.
This is because \citet{OhyamaEtAl_2007e} determined the dispersion and the wavelength origin in equation (\ref{eq. tool}) so that the equation represented the positions of the emission lines, but they did not consider whether it predicted the incident positions of the second-order light or not.
The one-sigma uncertainty of the revised wavelength calibration estimated from using $L$ and $\alpha$ is $\pm 0.1\,\mathrm{pix}$ or less (figure \ref{fig. wavelength_vs_pos}(c)).

Taking the refractive index as a function of the wavelength, we can explain the observed difference of the previous wavelength calibration.
For the first time, to the best of our knowledge, it is also shown that, owing to the wavelength dependence of the refractive index, the contamination by the second-order light occurs even in the case of the perfect order-sorting filter.

\subsection{Spectral Response Calibration}
\label{subsec. res-cal}

We obtained the relation between pixel offset and wavelength in the previous section.
Next, we discuss the conversion of the output of the detector into the flux density, or flux calibration.
We consider the response of the system not only for the first-order light but also for the second-order one to quantify and subtract the second-order contamination.
The method to obtain the responses is already explained by \citet{BabaEtAl_2016}.
Here, we describe it in more detail.

Raw spectroscopic images are provided in analog-to-digital units (ADU).
A series of basic data reduction procedures (dark subtraction, linearity correction, flat fielding, stacking, etc.) and spectral image extraction are performed using the toolkit \citep{OhyamaEtAl_2007e}.

If there is no contamination from the second-order light, the output at each pixel $N$ in ADU is proportional to the flux density $F_\nu(\lambda)$.
Hence, $N=R(\lambda)F_\nu(\lambda)$, where $R(\lambda)$ is the total spectral response of the system at wavelength $\lambda$.
Even if contamination by the second-order light occurs, similar relations hold for the individual components of the first- and the second-order light, respectively.
Therefore, the output at the $i$-th pixel $N_i$ can be written as
\begin{equation}
  N_i = R^{(1)}(\lambda^{(1)}_i) F_\nu(\lambda^{(1)}_i) + R^{(2)}(\lambda^{(2)}_i) F_\nu(\lambda^{(2)}_i).
  \label{eq. ADU}
\end{equation}
Here, $\lambda^{(1)}_i$ and $\lambda^{(2)}_i$ are the wavelengths of the first- and the second-order light in the $i$-th pixel.
$R^{(1)}(\lambda)$ and $R^{(2)}(\lambda)$ are the response functions from the first- and the second-order light.
Note that the two functions are not equal ($R^{(1)}(\lambda) \neq R^{(2)}(\lambda)$) because the grism does not disperse higher order light with the same efficiency as the first-order.

The present response $R^{(1)}(\lambda)$ from the first-order light is derived from observations of  A- and K-type standard stars \citep{OhyamaEtAl_2007e}.
Since these types of stars show blue spectra (Rayleigh-Jeans side) in 2.5--5.0\,\micron, this calibration scheme may result in severe contamination by second-order light, which leads to overestimates of $R^{(1)}$.

Because equation (\ref{eq. ADU}) contains two unknown responses $R^{(1)}$ and $R^{(2)}$, it cannot be solved with one standard star.
Even if the equation is simultaneously applied to two different standard stars, the response from the second-order light $R^{(2)}$ will have large uncertainty because these standard stars have similar spectra and do not provide sufficiently independent information for calibration.
To reliably obtain $R^{(1)}(\lambda)$ and $R^{(2)}(\lambda)$, we use standard objects that have blue and red spectra, where the latter suffer much less from the second-order light than the ordinary standard stars do.

\begin{figure}
  \begin{center}
    \includegraphics[width=16cm]{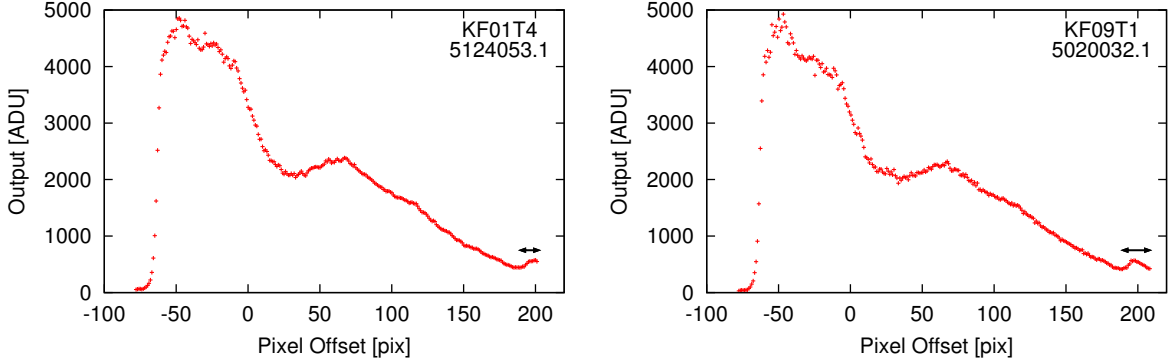}
  \end{center}
  \caption{Raw spectra of standard stars observed in the grism mode. The arrows denote the ranges where the second-order light component causes anomalies. Left: KF01T4 (observation ID: 5124053.1). Right: KF09T1 (observation ID: 5020032.1).}
  \label{fig. spec_stars}
\end{figure}

As for the blue standard objects, we use ordinary K-type standard stars.
Table \ref{tab. obs_stars} summarizes Two-Micron All-Sky Survey (2MASS; \cite{SkrutskieEtAl_2006}) IDs, spectral types, and observational information of the two standard stars.
KF09T1 is also used in the original spectral response calibration of the grism mode \citep{OhyamaEtAl_2007e}.
Using the toolkit, we extract the raw spectra.
Only the sky fluctuation of the spectral image is taken as the uncertainty of the output.
The obtained raw spectra of the stars are shown in figure \ref{fig. spec_stars}.
We use model spectra provided by M.~Cohen and coworkers (\cite{CohenEtAl_1996f}, \yearcite{CohenEtAl_1999d}, \yearcite{CohenEtAl_2003a}, \yearcite{CohenEtAl_2003}; \cite{Cohen_2003}) in the same manner as previous calibrations of the IRC (\cite{OhyamaEtAl_2007e}, \cite{TanabeEtAl_2008}, \cite{ShimonishiEtAl_2013}).
Both raw spectra show small bumps at around 200\,pix.
We attribute these bumps to the contaminating second-order light.

\begin{table}
  \tbl{Basic properties and observation log of standard stars.}{%
    \begin{tabular}{cccccc}
      \hline
      Object & 2MASS ID            & Type     & Obs. ID   & Obs. Date   & aperture \\
      \hline
      KF01T4 & J18040314$+$6654459 & K1.5 III & 5124053.1 & 2007 Apr  7 & Nc       \\
      KF09T1 & J17592304$+$6602561 & K0 III   & 5020032.1 & 2006 Apr 24 & Nc       \\
      \hline
    \end{tabular}
  }
  \label{tab. obs_stars}
\end{table}

\begin{figure}
  \begin{center}
    \includegraphics[width=16cm]{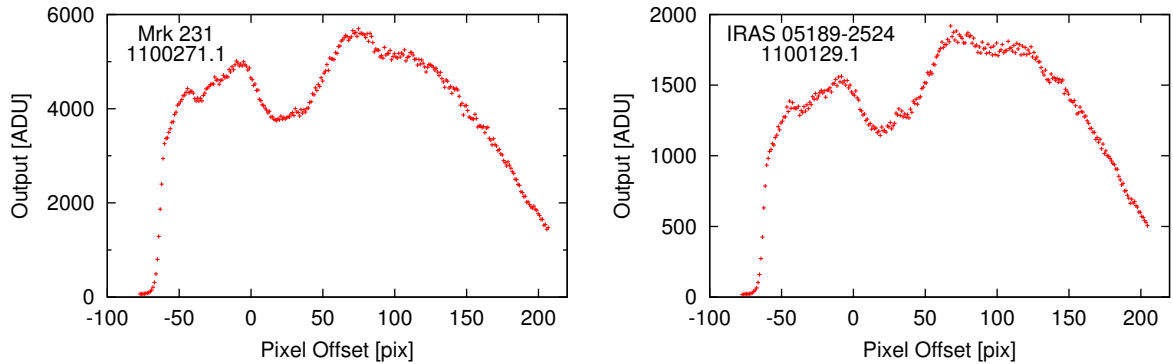}
  \end{center}
  \caption{Raw spectra of ULIRGs observed in the grism mode. Left: Mrk 231 (observation ID: 110027.1). Right: IRAS 05189$-$2524 (observation ID: 1100129.1).}
  \label{fig. spec_ULIRGs}
\end{figure}

We use two ultra-luminous infrared galaxies (ULIRGs) Mrk 231 and IRAS 05189$-$2524 as the red standard objects.
The energy sources of these ULIRGs are dominated by AGNs rather than starburst activity \citep{ImanishiEtAl_2000f}.
Owing to the strong thermal radiation from the dust heated by AGNs, the spectra of the two ULIRGs do not show measurable emission or absorption features in the 3--4\,\micron\ range except for the 3.3\,\micron\ polycyclic aromatic hydrogen (PAH) emission \citep{ImanishiEtAl_2000f, ImanishiEtAl_2007d}.
Although IRAS 05189$-$2524 shows weak 3.4\,\micron\ absorption of carbonaceous dust with optical depth $\sim 0.04$ \citep{ImanishiEtAl_2000f}, it is negligible in building the model spectrum.
The 3.4\,\micron\ absorption would affect the continuum by less than 4\%.
Table \ref{tab. obs_ULIRGs} summarizes the redshift, optical classification, and observational information of the two ULIRGs.
The raw spectra of the two ULIRGs are extracted in the same manner as the stars and shown in figure \ref{fig. spec_ULIRGs}.

\begin{table}
  \tbl{Basic properties and observation log of ULIRGs.}{%
    \begin{tabular}{cccccc}
      \hline
      Object            & redshift & Class\footnotemark[$*$] & Obs. ID   & Obs. Date     & aperture \\
      \hline
      Mrk 231           & 0.042    & Seyfert 1               & 1100271.1 & 2007 May 30   & Np       \\
      IRAS 05189$-$2524 & 0.043    & Seyfert 2               & 1100129.1 & 2007 Mar  8   & Np       \\
      \hline
    \end{tabular}
  }
  \label{tab. obs_ULIRGs}
  \begin{tabnote}
    \footnotemark[$*$] \citet{VeilleuxEtAl_1999a}.
  \end{tabnote}
\end{table}

To create the intrinsic model spectra of the ULIRGs, we compile observations made with telescopes other than AKARI.

Archival data of the 2MASS, Wide-field Infrared Survey Explorer (WISE; \cite{WrightEtAl_2010a}), and Spitzer/IRS are used.
From the 2MASS All-Sky Extended Source Catalog, we take the $J$, $H$, and $K_s$ total magnitudes extrapolated from the surface brightness profile \citep{SkrutskieEtAl_2006}.
The magnitudes are converted into flux densities using the mean wavelengths and corresponding flux densities for the zero magnitude provided by \citet{RiekeEtAl_2008}.
Under the definition of the mean wavelength and zero magnitude, the color correction is normalized for the spectral energy distribution of $F_\lambda=\mathrm{constant}$, or $F_\nu \propto \lambda^2$ (see Appendix E of \cite{RiekeEtAl_2008}).
Since the two ULIRGs show the same spectral shape in 1--2\,\micron, no color correction factor is applied.

\begin{figure}
  \begin{center}
    \includegraphics[width=16cm]{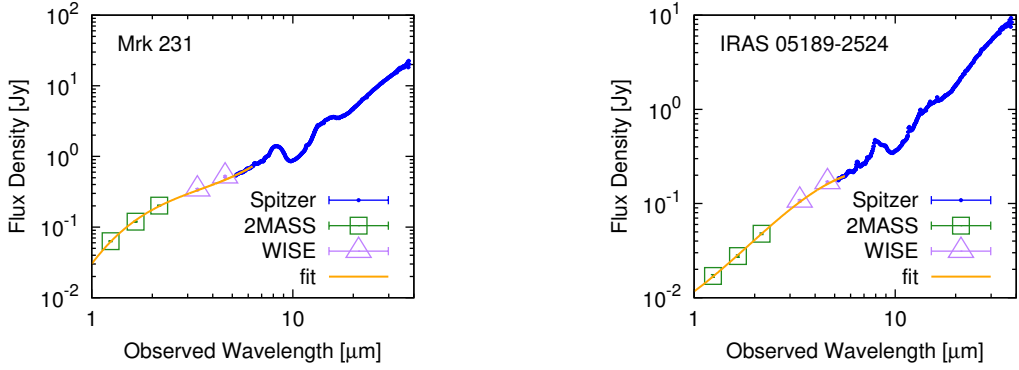}
  \end{center}
  \caption{Flux densities of the ULIRGs observed with 2MASS, WISE, and Spitzer/IRS. Solid lines denote the fitted cubic functions, which are taken as the model spectra of the ULIRGs. The fitting ranges are intervals where the lines are drawn. Left: Mrk 231. Right: IRAS 05189$-$2524.}
  \label{fig. spec_models}
\end{figure}

The W1 and W2 profile-fit magnitudes are taken from the AllWISE Source Catalog.
The fluxes for the zero magnitudes reported by \citet{JarrettEtAl_2011a} and the color correction factors provided by \citet{WrightEtAl_2010a} are used.
The calibrated Spitzer/IRS spectroscopic data are also considered.
These data are plotted in figure \ref{fig. spec_models}.
We fit a cubic function to these data points for each ULIRG in the $\log F_\nu$-$\log \lambda$ plane to approximate their spectra.
The fitted functions are shown in figure \ref{fig. spec_models}.

\begin{figure}
  \begin{center}
    \includegraphics[width=16cm]{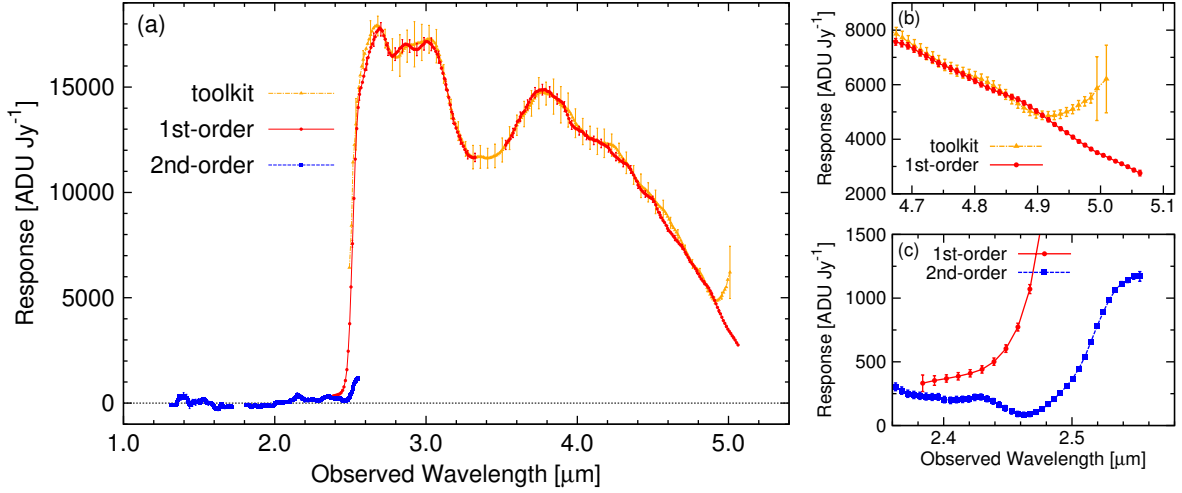}
  \end{center}
  \caption{Spectral responses from the first- and second-order light and that of the present toolkit. (a) The entire plot. Gaps are masked ranges affected by the PAH emission. Error bars are drawn every five points. (b) and (c) Zoom-in plots around 5\,\micron\ and 2.5\,\micron, respectively. The scales of the abscissas of panels (b) and (c) are aligned so that the same pixel offsets line up vertically (e.g., $\lambda^{(2)}=2.50\,\micron$ and $\lambda^{(1)}=4.95\,\micron$).}
  \label{fig. responses}
\end{figure}

We calculate the spectral responses from two pairs of standard objects (KF01T4 and Mrk 231, and KF09T1 and IRAS 05189$-$2524).
For the uncertainties of the responses, those in the raw spectra, model spectra, and the wavelength calibration curve are propagated.
The response from the first-order light has a measurement uncertainty of about 5\%.
The measurement uncertainty of the response from the second-order light is about 5\% at 2.55\,\micron\ and increases at shorter wavelengths.
At 2.3\,\micron, it becomes around 50\%.
Below 2.3\,\micron, the response from the second-order light is substantially zero.
The responses have systematic uncertainty of at most 4\% owing to the weak emission or absorption lines of the ULIRGs.
Since we built the model spectra of the ULIRGs ignoring the PAH emission, we mask the data points affected by the PAH emission.
The rest-frame full width at half-maximum of the PAH emission is 0.05\,\micron\ \citep{ImanishiEtAl_2000f, ImanishiEtAl_2007d}; thus, we remove 15 data points ($\sim 0.15\,\micron$) of the responses from the first-order light around $\lambda^{(1)}=(1+z)\times 3.29\,\micron$, where $z$ is the redshift of the ULIRGs and 3.29\,\micron\ is the peak wavelength of the PAH emission \citep{ImanishiEtAl_2008c}.
The data points of the responses from the second-order light are also removed at wavelengths that correspond to the same masked pixel offsets.
The two ULIRGs have similar redshifts, as listed in table \ref{tab. obs_ULIRGs}.
Hence, the PAH-masked regions almost entirely overlap with each other.
The results from the two pairs are averaged to reduce the uncertainty.
Next, since the responses hardly change along a few pixels, we take the five-pixel-width moving average to reduce the pixel-to-pixel scatter in a manner similar to the spectral response calibration for the prism mode by \citet{ShimonishiEtAl_2013}.
Finally, the responses shown in figure \ref{fig. responses} are obtained.
The set of the responses ($R^{(1)}$ and $R^{(2)}$) covers the range of $\Delta Y$ from $-$77.2 to 200.8\,pix.
This range corresponds to that of $\lambda^{(1)}=2.38$--$5.06\,\micron$ and $\lambda^{(2)}=1.31$--$2.55\,\micron$.
The measurement uncertainty of the response from the first- and second-order light is minimized to about 2\% and 2\%--26\%, respectively.

The obtained spectral response from the first-order light $R^{(1)}$ agrees with the previous one within the uncertainties between 2.6 and 4.9\,\micron, where contamination by the second-order light is not expected.
At wavelengths longer than 4.9\,\micron, $R^{(1)}$ of this work monotonically decreases in contrast to that in the toolkit.
The ratio of $R^{(1)}$ of this work to that of the toolkit is 0.63 at 5.0\,\micron\ and 0.96 at 2.6\,\micron.
In addition to this, the response from the second-order light $R^{(2)}$, which is, to our knowledge, derived for the first time, increases at wavelengths longer than 2.5\,\micron. 
These results show that the previous response curve at wavelengths longer than 4.9\,\micron\ is contributed by the two components, first- and second-order light, and that we succeed in separating them into the two responses $R^{(1)}$ and $R^{(2)}$.
Moreover, the responses are significantly nonzero at wavelengths shorter than 2.5\,\micron.
This means that the detector responds to the radiation of wavelengths shorter than 2.5\,\micron, beyond the nominal wavelength range of 2.5--5.0\,\micron.
Therefore, this suggests that the order-sorting filter has some leakage below 2.5\,\micron.

Subtraction of the second-order-light component from the first-order spectra is formalized by extending the spectral response in the matrix that relates the output with the first-order spectrum.
The first-order spectrum can be purely obtained from this equation. 
This is the main goal of our calibrations.

To evaluate the amount of the second-order light as a function of the wavelength, we need the intensity of the first-order light at the same wavelength.
We assume that wavelength $\lambda^{(1)}_1$, which comes in at the first pixel as first-order light, enters between the $(k-1)$-th and $k$-th pixels as second-order light.
Based on this assumption, since the increment of $\lambda^{(2)}_i$ is about half of that of $\lambda^{(1)}_i$, the following magnitude relation holds:
\begin{equation}
  \lambda^{(2)}_{k-1} < \lambda^{(1)}_1 < \lambda^{(2)}_k < \lambda^{(2)}_{k+1} < \lambda^{(1)}_2 < \cdots.
  \label{eq. increments}
\end{equation}
Hereafter, we define
$ R^{(1)}(\lambda^{(1)}_{i})$ as $R^{(1)}_{i}$, $ R^{(2)}(\lambda^{(2)}_{i})$ as $R^{(2)}_{i}$, and $ F_\nu(\lambda^{(1)}_{i})$ as $F_{\nu,i}$.
Using these notations, the output at the $k$-th pixel is
\begin{equation}
  N_k = R^{(1)}_k F_{\nu,k} + R^{(2)}_k F_\nu(\lambda^{(2)}_k).
  \label{eq. ADU_k}
\end{equation}
From the linear interpolation, the flux density of the second-order-light component at the $k$th pixel becomes
\begin{equation}
  F_\nu(\lambda^{(2)}_k) = \frac{\lambda^{(1)}_2-\lambda^{(2)}_k}{\lambda^{(1)}_2-\lambda^{(1)}_1} F_{\nu,1} + \frac{\lambda^{(2)}_k-\lambda^{(1)}_1}{\lambda^{(1)}_2-\lambda^{(1)}_1} F_{\nu,2}.
  \label{eq. linterp}
\end{equation}
Equations (\ref{eq. ADU_k}) and (\ref{eq. linterp}) can be combined into the following matrix.
{\setlength\arraycolsep{1pt}
\begin{equation}
  \left(
  \begin{array}{c}
    N_1    \\
    N_2    \\
    \vdots \\
    N_k    \\
    \vdots \\
  \end{array}
  \right)
  =
  \left(
  \begin{array}{llllll}
    R_{1,1}                        &                                &        &               &              \\
                                   & R_{2,2}                        &        &               &              \\
                                   &                                & \ddots &               &              \\
    R_{k,1}                        & R_{k,2}                        &        & \; R_{k,k} \; &              \\
    \quad\rotatebox{-25}{$\ddots$} & \quad\rotatebox{-25}{$\ddots$} &        &               & \ddots \quad \\
  \end{array}
  \right)
  \left(
  \begin{array}{c}
    F_{\nu,1} \\
    F_{\nu,2} \\
    \vdots    \\
    F_{\nu,k} \\
    \vdots    \\
  \end{array}
  \right)
  \label{eq. res_mtrx}
\end{equation}}
\begin{eqnarray}
  && R_{i,i} = R^{(1)}_i \label{eq. diag} \\
  && R_{k,1} = \frac{\lambda^{(1)}_2 - \lambda^{(2)}_k }{\lambda^{(1)}_2 - \lambda^{(1)}_1} R^{(2)}_k , \,
     R_{k,2} = \frac{\lambda^{(2)}_k - \lambda^{(1)}_1 }{\lambda^{(1)}_2 - \lambda^{(1)}_1} R^{(2)}_k , \, \cdots \label{eq. off-diag}
\end{eqnarray}
The response from the second-order light is included as the off-diagonal elements of this matrix.
The inverse matrix of the response matrix can be analytically obtained (see Appendix).
Multiplying it with the column vector of the output, we can obtain the pure first-order spectrum $F_{\nu,i}$.

The spectral responses obtained in this section separate the components of the first- and second-order light and quantify the second-order-light contamination.
The relations among the components of the first- and second-order light and the output can be formalized in one equation with the response matrix, whose diagonal and off-diagonal elements represent the responses from the first- and second-order light, respectively.
The flux calibration that considers the contamination effect can be achieved by multiplying the inverse matrix of the response with the output.

\section{Demonstrations of the Effectiveness of the New Flux Calibration}
\label{sec. flux-cal}

We here demonstrate the matrix-formulated flux calibration for two objects that have different colors and compare the results with those from the toolkit.
In the following demonstrations, the response from the first-order light is interpolated into the PAH-masked range using a quadratic function of the wavelength.

\begin{figure}
  \begin{center}
    \includegraphics[width=8cm]{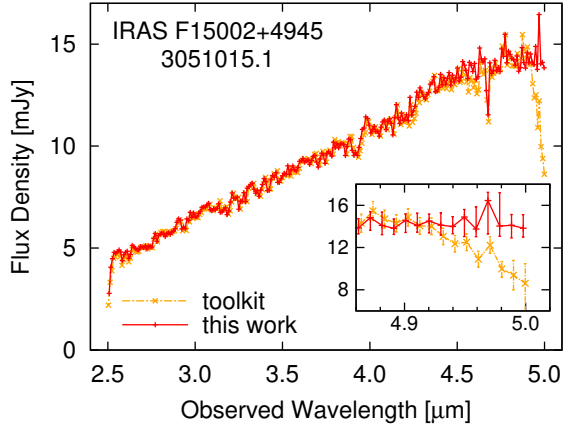}
  \end{center}
  \caption{Demonstration of the matrix-formulated calibration for an ULIRG, IRAS F15002+4945 (observation ID: 3051015.1). The spectrum processed with the toolkit is also shown.}
  \label{fig. demo_red}
\end{figure}

Figure \ref{fig. demo_red} shows the spectrum of an ULIRG (IRAS F15002+4945) with the new calibration matrix.
This ULIRG has a red spectrum, and hence, the component of the second-order light is expected to be relatively smaller than that of the blue objects.
The spectrum calculated using the new flux calibration shows a smooth distribution up to 5.0\,\micron, whereas that obtained by the toolkit decreases from 4.9\,\micron\ to 5.0\,\micron.
This difference stems from that of the current and previous responses from the first-order light.
The current response monotonically decreases at around 4.9\,\micron, whereas the previous one increases at the same point (see figure \ref{fig. responses}(b)).

\begin{figure}
  \begin{center}
    \includegraphics[width=8cm]{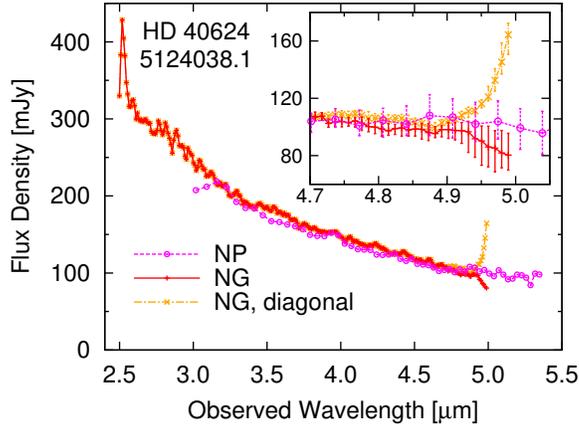}
  \end{center}
  \caption{Demonstration of the matrix-formulated calibration for an A0V star, HD 40624 (observation ID: 5124038.1). The spectrum calculated with only the diagonal elements of the response matrix is also shown. The magenta points show the spectrum obtained in the prism mode observation of IRC.  The prism mode spectrum in the 1.8--3.0\,\micron\ range was not obtained owing to the saturation of the detector.}
  \label{fig. demo_blue}
\end{figure}

As another example, figure \ref{fig. demo_blue} shows the result for an A0V star (HD 40624), which has a blue spectrum.
In this case, the first-order spectrum at longer wavelengths is assumed to be largely contaminated by the second-order light that comes from shorter wavelengths.
This star was also observed in the prism mode just before the grism mode observation.
Since the prism is not a grating element, any problems associated with the higher order light cannot occur in the prism mode.
The flux calibration results basically agree with the prism mode spectrum but disagree with it if the off-diagonal elements of the response matrix, which measure the contribution of the second-order light, are ignored.
This suggests that our flux calibration successfully removes the second-order component.

The toolkit has not been able to correct the second-order light contamination of the first-order spectra in wavelengths longer than 4.9\,\micron.
This biases the flux calibration, especially for red objects, and prevents obtaining accurate spectra in this wavelength range.
The new flux calibration, which quantifies and subtracts the contamination, can resolve this problem for a wide range of objects, as shown by the above examples.
Now we can accurately obtain the 4.9--5.0\,\micron\ spectra without the effect of the second-order light.
This allows the study of the spectral features of red objects that appear in this wavelength range, such as the carbon monoxide absorption within nearby AGNs from the AKARI grism mode spectroscopic observations.
\citet{ImanishiEtAl_2008c} reported that three ULIRGs observed in Phases 1 and 2 by the grism mode show the CO absorption feature.
However, these absorption spectra are significantly affected by the second-order contamination and have never been analyzed to date. 
Moreover, OCS ice absorption (4.9\,\micron) falls into the extended wavelength range \citep{BoogertEtAl_2015}.
The OCS ice can be used to study the evolution of circumstellar disks of young stellar objects.
For instance, \citet{AikawaEtAl_2012} tentatively detected the OCS ice absorption toward a low-mass young stellar object IRC-L1041-2 from the grism mode observation.
To study this feature in young stellar objects, the effect of the contamination needs to be corrected.

\section{Summary}
\label{sec. summary}

In this paper, we revise the calibrations of the AKARI IRC grism mode spectroscopy by subtracting the second-order light that contaminates the first-order spectrum.
First, the refractive index of Ge, the material of which the grism is made, is taken as a function of the wavelength.
The difference from the previous linear wavelength calibration is measured from the recombination lines of planetary nebulae, and we successfully explain it by the wavelength dependence of the refractive index.
Next, the spectral responses from the first- and second-order light are simultaneously obtained with standard objects that show contrastive red and blue spectra.
With these new responses, the first- and second-order light mixing in 4.9--5.0\,\micron\ are decomposed for the first time.
Finally, the flux calibration with the set of the responses is formulated into a matrix form.
This flux calibration demonstrates and succeeds in removing the second-order-light component in both red and blue objects.
The new calibrations enable us to obtain correct 4.9--5.0\,\micron\ spectra from grism mode observations and are useful in studies of, for example, the carbon monoxide absorption within nearby AGNs.
Note that the calibrations presented here are limited to the observations before the exhaustion of liquid helium (Phases 1 and 2).
During the post-helium phase (Phase 3), the temperature of the detector exceeds 40\,K, and hence, the operating conditions in Phase 3 largely differ from those in Phases 1 and 2.
Similar calibrations for Phase 3 will be presented in forthcoming papers.

\begin{ack}
AKARI is a JAXA project with the participation of ESA.
We thank all the members of the AKARI project for their continuous help and support.
We are grateful to H.~Matsuhara and T.~Wada for their expertise on the IRC.
We thank the anonymous referee for careful reading our manuscript and for helpful comments that improved the quality of this paper.
This publication makes use of data products from the Two Micron All Sky Survey, which is a joint project of the University of Massachusetts and the Infrared Processing and Analysis Center/California Institute of Technology, funded by the National Aeronautics and Space Administration and the National Science Foundation.
This publication makes use of data products from the Wide-field Infrared Survey Explorer, which is a joint project of the University of California, Los Angeles, and the Jet Propulsion Laboratory/California Institute of Technology, funded by the National Aeronautics and Space Administration.
This work is based in part on observations made with the Spitzer Space Telescope, which is operated by the Jet Propulsion Laboratory, California Institute of Technology under a contract with NASA.
This study was supported by JSPS KAKENHI Grant Number 26247030.
Y.O. acknowledges support from MOST grant 104-2112-M-001-034-.
\end{ack}

\appendix 
\section*{Inverse Matrix of the Response Matrix}

\begin{figure}
  \begin{center}
    \includegraphics[width=8cm]{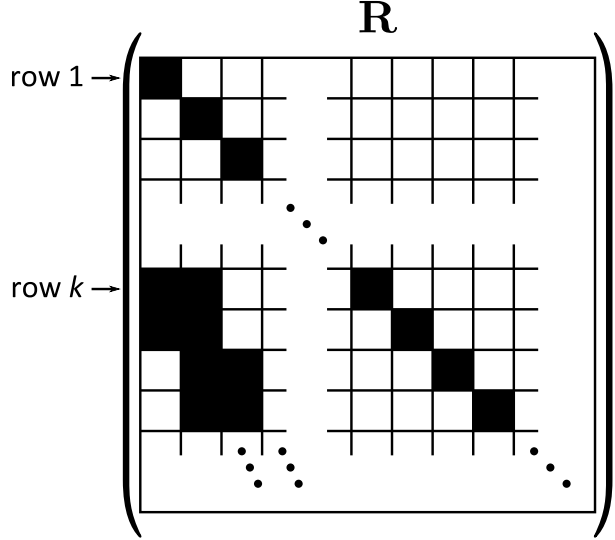}
  \end{center}
  \caption{Elements of the response matrix $\mathbf{R}$. Black cells denote the nonzero elements. All other elements are zero.}
  \label{fig. nonzero}
\end{figure}

We denote the response matrix defined by equations (\ref{eq. res_mtrx}), (\ref{eq. diag}), and (\ref{eq. off-diag}) as $\mathbf{R}$.
$\mathbf{R}$ is a triangular matrix.
Figure \ref{fig. nonzero} displays the nonzero elements of $\mathbf{R}$.
The columns with nonzero off-diagonal elements shift rightward every two rows because the magnitude relation (\ref{eq. increments}) holds.

By the sweep-out method or similar, the inverse matrix $\mathbf{D} \equiv \mathbf{R}^{-1}$ is obtained as follows:
\begin{equation}
  \mathbf{D}=
  \left(
  \begin{array}{llllll}
    D_{1,1}                        &                                &        &               &              \\
                                   & D_{2,2}                        &        &               &              \\
                                   &                                & \ddots &               &              \\
    D_{k,1}                        & D_{k,2}                        &        & \; D_{k,k} \; &              \\
    \quad\rotatebox{-25}{$\ddots$} & \quad\rotatebox{-25}{$\ddots$} &        &               & \ddots \quad \\
  \end{array}
  \right)
\end{equation}
\begin{eqnarray}
  && D_{i,i} = \frac{1}{R_{i,i}}               \\
  && D_{i,j} = -\frac{R_{i,j}}{R_{i,i}R_{j,j}}
\end{eqnarray}


\end{document}